\documentclass[aps,floats,epsf, twocolumn]{revtex4}
 \usepackage{graphics}

\begin{document}

\title{Topological insulator in the core of the superconducting vortex in graphene}
\author{Igor F. Herbut}

\affiliation{Department of Physics, Simon Fraser University,
 Burnaby, British Columbia, Canada V5A 1S6}

\begin{abstract}
The core of the vortex in a general superconducting order parameter in graphene is argued to be ordered, with the possible local order parameters forming the algebra $U(1)\times Cl(3)$. A sufficiently strong Zeeman coupling of the magnetic field of the vortex to the electron spin breaks the degeneracy in the core in favor of the anomalous quantum Hall state. I consider a variety of superconducting condensates on the honeycomb lattice and demonstrate the surprising universality of this result. A way to experimentally determine the outcome of the possible competition between different types of orders in the core is proposed.
\end{abstract}
\maketitle

\vspace{10pt}

  Besides the usual metallic state \cite{geim}, Dirac quasiparticles in graphene may, at least in principle, form a multitude of insulating phases \cite{herb-jur-roy}. The insulators should be understood as different types of dynamically or externally generated masses of Dirac fermions, which preserve the quasi-relativistic invariance of the low-energy theory, but break some of the space-time symmetries of the Hamiltonian \cite{dima}. Yet another type of ordered phase of electrons on honeycomb lattice is the superconductor,  which breaks the U(1) symmetry associated with the conservation of the particle number \cite{dima, zhao, honerkamp, bergman}. The superconducting state should be inducible in graphene by the proximity effect for example \cite{beenaker}. At the level of the continuum Dirac Hamiltonian,  superconducting states in graphene may be distinguished by their symmetry with respect to the inversion of the two inequivalent Dirac points, as the s and p waves, and with respect to the translational invariance,  as  uniform and not. The variety of such superconducting states notwithstanding, I show here that the core of the superconducting vortex in undoped graphene is always ordered, and that, at least in the simplest model, always in the same way: the core of the vortex is the Haldane-Kane-Melle (HKM) topological insulator \cite{haldane, kane}.

  The above result is obtained within two complementary Dirac representations of the Bogoliubov-de Gennes  (BdG) quasiparticle Hamiltonian, in which it arises in related, but distinct ways. The problem of the internal structure  of the superconducting vortex in graphene may be mapped onto the general Dirac Hamiltonian in presence of two-component mass-term with a twist. It is well known that the single zero-energy state in this situation would imply a quantum number fractionalization \cite{jackiw}. In graphene, however, the spin-1/2 of the electrons dictates two copies of Dirac fermions, and the concomitant doubling of the number of zero-energy states restores the usual quantization rule. The same zero-energy states, nevertheless, are responsible for another nontrivial property of the vortex core: it always harbors a finite local order parameter (OP) of some kind. An early example of this effect was provided in ref. \cite{herbutAF}, where the vortex configuration of the in-plane N\' eel OP on honeycomb lattice was shown to induce an out-of-plane component of the same OP. More recently, the superconducting vortex was argued to exhibit charge-density waves (CDW) and bond-density waves (BDW) in the core \cite{ghaemi}. Here it will be shown that in general the vortex of unit vorticity in the mass of the two-copy four-component Dirac Hamiltonian allows {\it four} order parameters, which close $U(1)\times Cl(3)$, where $Cl(3)$ stands for the three-dimensional Clifford algebra. The members of the algebra depend on the type of the underlying order; nevertheless, in a superconductor of arbitrary symmetry, if the magnetic field in the vortex core is sufficiently strong the HKM anomalous quantum Hall state is {\it universally} the ground state. The effect is due to the Zeeman coupling of the magnetic field to the electron spin, which, quite unexpectedly, always selects the above state out of a wide variety of competing possibilities.

   The HKM insulator is interesting in its own right, due to its quantized spin Hall effect, and the accompanying  structure of the edge states \cite{kane}. Such a topologically nontrivial ground state is favored by the spin-orbit interactions in graphene, which however are well known to be way too weak to lead to an observable effect. It was recently proposed that this ground state may also result from the electron-electron Coulomb repulsion in the presence of a bulge in the graphene sheet  \cite{herbut-catalysis}. The present considerations suggest that placing graphene  on top of a type-II superconductor in a mixed state may provide an alternative, albeit local, realization of this elusive state of matter.

  The announced result will be first derived within the rotationally invariant representation of the paring Hamiltonian. Graphene exhibits gapless excitations near two inequivalent Dirac points at $\pm \vec{K}$, with $\vec{K}= (1,1/\sqrt{3}) (2 \pi/a\sqrt{3})$, for example,  where $a$ is the lattice spacing \cite{herb-jur-roy}. One may form an {\it eight-component} Dirac-Nambu  fermion as
  $\Psi^\dagger = (\Psi_+ ^\dagger, \Psi_- ^\dagger)$, where
  \begin{widetext}
  \begin{equation}
  \Psi^\dagger _\sigma (\vec{k}, \omega) =
  ( u^\dagger _\sigma (\vec{k}, \omega), v^\dagger _\sigma (\vec{k}, \omega), \sigma u_{-\sigma} (-\vec{k}, -\omega), \sigma v_{-\sigma} (-\vec{k}, -\omega) ),
  \end{equation}
  \end{widetext}
  where $\sigma=\pm$ labels the projection of the electron spin along the z-axis, and $\vec{k}= \vec{K}+ \vec{p}$, with $|\vec{p}| \ll K$. $u_\sigma $ and $v_\sigma $ are the Grassmann variables corresponding to two triangular sublattices of the honeycomb lattice. The imaginary-time Lagrangian for the  excitations near the Dirac points in this representation becomes
  \begin{equation}
  L =  \Psi ^\dagger (\vec{x},\tau ) (\partial_\tau + H_0 )  \Psi (\vec{x},\tau),
  \end{equation}
  where the single-particle Hamiltonian is  $H_0 = I_2 \otimes i\gamma_0 \gamma_i \hat{p}_i $, with
   $\gamma_0= \sigma_3 \otimes \sigma_3$, $\gamma_1 = I_2 \otimes \sigma_2$, and  $\gamma_2 = I_2 \otimes \sigma_1$, and $i=1,2$. We may then choose $\gamma_3= \sigma_1 \otimes \sigma_3$  and  $\gamma_5 = \sigma_2 \otimes \sigma_3$. $\{ I_2, \vec{\sigma} \}$ is the Pauli basis of two-component matrices.

  The Lagrangian is invariant under the global  $U(4)$ symmetry generated by the 16 matrices
  $\{ I_2, \vec{\sigma} \} \otimes \{ I_4, \gamma_3, \gamma_5, \gamma_{35} \} $, where $\gamma_{35}= i\gamma_3\gamma_5 = \sigma_3 \otimes I_2$, and $I_4$ is the four-dimensional unit matrix. The $SU(2)$  subgroup generated by the $\vec{S}= \vec{\sigma} \otimes I_4$ is the group of rotations of the electron spin, and is, of course, exact. This is due to the factor ``$\sigma$" in the right half of the Dirac-Nambu fermion in Eq. (1), which ensures that the rotation may be accomplished by the matrix multiplication from the left. The above representation is in this sense {\it manifestly} rotationally invariant. It will also prove helpful to recognize $N = I_2 \otimes \gamma_{35}$ as the number operator.

  There are also 16 linearly independent Hermitian matrices that {\it anticommute}  with $H_0 $: $\{ I_2, \vec{\sigma} \} \otimes \{ \gamma_0, i\gamma_0 \gamma_3, i\gamma_0 \gamma_5, i\gamma_1\gamma_2 \}$. The addition of any of these matrices to the Hamiltonian $H_0$  would gap the Dirac spectrum.
  The matrices that correspond to a superconducting orders are those that {\it do not commute} with the particle-number. It is easy to see that
    $ \langle \Psi^\dagger ( I_2 \otimes i \gamma_0 \gamma_3)  \Psi \rangle = Re \Delta_s$, and
     $ \langle \Psi^\dagger ( I_2 \otimes i \gamma_0 \gamma_5)  \Psi \rangle = Im \Delta_s$, where $\Delta _s $ represents the spatially uniform, complex s-wave superconducting OP. Consider then the BdG pairing Hamiltonian in the presence of the vortex in the underlying s-wave superconducting ground state:
     \begin{equation}
     H_{BdG} = H_0 + |\Delta_s (r)|  I_2 \otimes [i\gamma_0 (\gamma_3\sin \theta+   \gamma_5 \cos \theta )],
     \end{equation}
     where $(r,\theta)$ are the polar coordinates in the graphene plane. $|\Delta_s (r\rightarrow \infty)|=const.$, and otherwise arbitrary. The magnetic field in the vortex will be included shortly.   An index theorem \cite{weinberg} guarantees that the spectrum of the block-diagonal Hamiltonian $H_{BdG} $ contains {\it two} states with zero energy: $\Psi_{0,1} ^\dagger =(\psi_0,0)$ and $\Psi_{0,2} ^\dagger  = (0,\psi_0)$, where $\psi_0$ is the rotationally invariant,  four-component bound state \cite{jackiw}. The Hilbert space $ {\cal H}_0$ spanned by these two zero-energy states is invariant under all operators that commute or {\it anticommute} with $H_{BdG}$ \cite{herbut-catalysis}, with four falling into the latter category: $ \{ I_2, \vec{ \sigma }  \} \otimes \gamma_0 $. These are at the center of our study.

On the other hand, the expectation value of any traceless single-particle operator $M$ that anticommutes with the Dirac Hamiltonian is easily seen to derive entirely from the zero-energy states \cite{herbutAF}. At $T=0$:
     \begin{equation}
\langle \Psi^\dagger  M   \Psi \rangle =  \frac{1}{2} ( \sum_{i, oc}-
\sum_{i, unoc}) \psi_{0,i} ^\dagger (\vec{x}) M \psi_{0,i} (\vec{x}),
\end{equation}
 where $\{ \psi_{0,i} (\vec{x}) \} $ is a basis in ${\cal H}_0$. Since ${\cal H}_0$ is two-dimensional, the four anticommuting operators from above  in ${\cal H}_0$ reduce to the familiar Pauli matrices $\{I_2, \vec{ \sigma } \}$. In full analogy to the spin-1/2 problem \cite{ballentine}, any two orthogonal states in ${\cal H}_0$ are then the $+1$ and $-1$ eigenstates of $(\hat{n} \cdot \vec{\sigma} ) \otimes \gamma_0$ for {\it some} unit vector $\hat{n}$ , and the $+1$ eigenstates of $I_2 \otimes \gamma_0$. Equation (4) then implies that: a) when one state is occupied and the second state empty,
  $\langle \Psi^\dagger ( \hat{n} \cdot \vec{\sigma} \otimes \gamma_0)  \Psi \rangle = \psi_0 ^\dagger \psi_0$, and $\langle \Psi^\dagger ( I_2 \otimes \gamma_0)  \Psi \rangle = 0$, and b) when both states are occupied or both empty  $\langle \Psi^\dagger ( I_2 \otimes \gamma_0)  \Psi \rangle = \pm \psi_0 ^\dagger \psi_0$, with  $\langle \Psi^\dagger ( \hat{n} \cdot \vec{\sigma} \otimes \gamma_0)  \Psi \rangle = 0$ for {\it all} $\hat{n}$.

  It is straightforward to rewrite these fermion bilinears in terms of the original electrons. For example
  \begin{equation}
  \Psi^\dagger ( I_2 \otimes \gamma_0)  \Psi = u_\sigma ^\dagger (\vec{x}, \tau) u_\sigma (\vec{x}, \tau ) - v^\dagger _\sigma (\vec{x}, \tau)  v_\sigma (\vec{x}, \tau).
 \end{equation}
 The corresponding average represents therefore the CDW. Similarly,
 \begin{eqnarray}
 \Psi^\dagger  (\vec{k},\omega) ( \sigma_3 \otimes \gamma_0)  \Psi (\vec{k},\omega) =
 \sigma  [ u_\sigma  ^\dagger (\vec{k},\omega ) u_\sigma  (\vec{k},\omega) \\ \nonumber
 - u_\sigma  ^\dagger (-\vec{k},-\omega ) u_\sigma  (-\vec{k},-\omega) ]  - [u\rightarrow v],
 \end{eqnarray}
 and may be recognized as the z-component of the vector OP for the HKM topological insulator \cite{haldane, kane}. The ground state with a finite expectation value of this operator would break the rotational invariance, the sublattice exchange symmetry, and, most importantly, the time reversal for each spin projection separately.

 Let us take the effect of the magnetic field of the vortex into account next. Assume the field to be along the z-direction, orthogonal to the graphene plane. First, the sole orbital effect of the localized magnetic field is the change of the form of the zero-energy states \cite{pi}. This follows from the observation that in the Dirac-Nambu representation the  magnetic field enters the Hamiltonian in Eq. (3) by modifying only the $H_0$ term as
 \begin{eqnarray}
 H_0 \rightarrow H_0 [A] = I_2 \otimes i\gamma_0 \gamma_i (\hat{p}_i - A_i \gamma_{35} ) = \\ \nonumber
 e^{-\chi(\vec{x}) I_2 \otimes \gamma_0}  H_0  e^{-\chi(\vec{x}) I_2 \otimes \gamma_0},
\end{eqnarray}
where the magnetic field is $B = \epsilon_{ij}\partial_i A_j = \partial^2 \chi$. (In our units $\hbar=e=c=1$.) Since the matrix $I_2 \otimes \gamma_0$ anticommutes with the mass term in Eq. (3), and acts as the unit operator within ${\cal H}_0$, the zero-energy states with and without the magnetic field differ only in the factor $\exp[\chi(\vec{x})]$. More important turns out to be the Zeeman effect on the electron spin. With the Zeeman term the Hamiltonian becomes $H_{BdG}+ H_Z$, where
\begin{equation}
H_Z = g B (\vec{x}) (\sigma_3 \otimes I_4),
\end{equation}
and $g\approx 2$ for the electrons in graphene.  One may consider $H_Z$ as a weak perturbation to the spectrum of $H_{BdG}$ \cite{wilczek}. The commutation relations with the four operators that anticommute with $H_{BdG}$ imply that, within ${\cal H}_0$,  $H_Z$ is proportional to $\sigma_3 \otimes \gamma_0$, and therefore it splits the  $+1$ and $-1$ eigenstates  of this particular OP. Equation (4) then implies that
 \begin{equation}
 \langle \Psi^\dagger ( \sigma_3 \otimes \gamma_0)  \Psi \rangle \propto
 |Exp[ \chi(\vec{x}) - \int_0 ^ r \Delta (r') dr'   ] |^2,
 \end{equation}
 whereas the averages of the other three OPs vanish. As the total flux of the magnetic field is $hc/2e$, we have $\chi(x) \approx (1/2) \ln|\vec{x}| $ at distances beyond the magnetic field's penetration depth, and the OP is exponentially localized within the coherence length $\xi=1/\Delta_s (r=\infty)$.

 Interestingly, the same conclusion follows even for the superconductor with a {\it different}  symmetry. Consider the BdG Hamiltonian for a vortex in the p-wave state:
\begin{equation}
     H_{BdG} = H_0 + |\Delta_p (r)|  \sigma_3 \otimes [ i\gamma_0  ( \gamma_3 \sin \theta +   \gamma_5 \cos \theta )],
\end{equation}
which breaks the spin-rotational symmetry. This superconducting OP is odd under the exchange of the Dirac points, and it is energetically preferable for strong  next-nearest neighbor attraction between electrons \cite{honerkamp}. The $U(1)\times Cl(3)$ algebra of OPs in the core is now different:  $U(1)=\{ I_2 \times \gamma_0 \}$ and $Cl(3)= \{ \sigma_3 \otimes \gamma_0, \sigma_1 \otimes i\gamma_1 \gamma_2,  \sigma_2 \otimes i\gamma_1 \gamma_2\}$. The last two matrices correspond to the x and y components of the N\' eel OP. The orbital effect of the magnetic field in resolving the degeneracy of the zero-energy states is still null. $H_Z$, on the other hand, within ${\cal H}_0$ is again proportional to $\sigma_3 \otimes \gamma_0$, as dictated by the commutation relation of the $H_Z$ with the above set of OPs. This implies the same result in Eq. (9) follows for the p-wave vortex as well.

Let us now construct a different representation of the problem in which the manifest rotational invariance of the Lagrangian in Eq. (2) is sacrificed \cite{herbutQED3}. Doing so will enable us to include the non-uniform insulating and superconducting states, and make contact with the recent work of Ghaemi et al. \cite{ghaemi}. Consider a {\it different} eight-component Dirac fermion  $\Phi^\dagger = (\Phi_+ ^\dagger, ( T  (\sigma_1 \otimes I_2) \Phi_+ ) ^\dagger)$, where the Dirac field $\Phi_+ $ is now rather standard \cite{herb-jur-roy}:
   \begin{widetext}
  \begin{equation}
  \Phi^\dagger _+ (\vec{q}, \omega) =
  ( u^\dagger _+ (\vec{K}+ \vec{q}, \omega), v^\dagger _+ (\vec{K}+ \vec{q}, \omega), u_+ ^\dagger (-\vec{K}+ \vec{q}, \omega), v_+ ^\dagger (-\vec{K}+ \vec{q}, \omega) ).
  \end{equation}
  \end{widetext}
 $T$ is the time-reversal operator, consisting of spin, momentum, and frequency reversal, and particle-hole operator exchange. The Lagrangian can still be written in the form in Eq. (2), but with the Hamiltonian $H_0$ replaced with $\tilde{H}_0 = (I_2 \otimes i\gamma_0 \gamma_1) \hat{p}_1 +  (\sigma_3 \otimes i\gamma_0 \gamma_2) \hat{p}_2$, and with $\gamma_0= I_2 \otimes \sigma_3$, $\gamma_1= \sigma_3 \otimes \sigma_2$, $\gamma_2= I_2 \otimes \sigma_1$, $\gamma_3= \sigma_1 \otimes \sigma_2$, and $\gamma_5= \sigma_2 \otimes \sigma_2$ \cite{herb-jur-roy}.

    Once the number operator in this representation is recognized as ${\tilde N} = \sigma_3 \otimes I_4$, the set of sixteen matrices that anticommute with $\tilde{H}_0$ may be divided into those corresponding to the superconducting
 ($\{ \sigma_1, \sigma_2 \} \otimes \{ \gamma_1, i\gamma_0 \gamma_2, i\gamma_1 \gamma_5, i \gamma_1 \gamma_3 \}$)
and the insulating OPs  ($\{ \sigma_3, I_2 \} \otimes \{ \gamma_0, i\gamma_0 \gamma_3, i\gamma_0 \gamma_5, i \gamma_1 \gamma_2 \}$). In revealing the orders corresponding to these fermion bilinears it is useful to discern $P = \sigma_3 \otimes \gamma_{35}$ as the generator of translations, and $I_s = I_2 \otimes i\gamma_1 \gamma_5 $ as the $\vec{K} \leftrightarrow -\vec{K}$ inversion operator \cite{herb-jur-roy}.  This immediately implies that the averages  $\langle \Phi^\dagger M \Phi \rangle $ represent: a) uniform s wave, for $M= \{ \sigma_1, \sigma_2 \} \otimes i\gamma_1 \gamma_5$, b) nonuniform s wave, for $M= \{ \sigma_1, \sigma_2 \} \otimes i\gamma_0 \gamma_2$, c) uniform p wave, for  $M= \{ \sigma_1, \sigma_2 \} \otimes i\gamma_1 \gamma_3$, and d) nonuniform p wave superconductor, for $M= \{ \sigma_1, \sigma_2 \} \otimes \gamma_1$. Explicit rewriting of the above bilinears shows that the non-uniform superconducting states have the $2\vec{K}$ periodicity of the OP, and thus represent the LOFF phases  \cite{loff} with the Kekule texture. Similarly, the insulating states are: a) CDW, for $M= \sigma_3  \otimes  \gamma_0$, b) spin-scalar  Kekule BDW \cite{hou}, for $M= \{ \sigma_3  \otimes i\gamma_0 \gamma_3, I_2 \otimes  i\gamma_0 \gamma_5 \}$, c) z-component of the HKM, for  $ M= I_2 \otimes i\gamma_1 \gamma_2$. Exchanging $\sigma_3$ and $I_2$ in the above further yields, a) N\' eel, b) spin-vector BDW, and c) spin-scalar Haldane's  OP, respectively.

  Let us reconsider the Hamiltonian with the vortex in the s-wave superconducting OP. In this representation it takes the form of
  \begin{equation}
   \tilde{H}_{BdG}= \tilde{H}_0 + |\Delta_s (r)| (\sigma_1 \sin \theta+  \sigma_2 \cos \theta ) \otimes  i\gamma_1 \gamma_5.
  \end{equation}
 The $U(1)\times Cl(3)$ OP algebra consists now of $Cl(3) = \{ \sigma_3 \otimes \gamma_0, \sigma_3 \otimes i\gamma_0 \gamma_3, I_2\otimes i\gamma_0 \gamma_5 \}$, and $U(1) = \{ I_2 \otimes i\gamma_1\gamma_2 \} $. The CDW and BDW belong to the $Cl(3)$ part, in agreement with the observation in ref. \cite{ghaemi}. In this representation the Zeeman term becomes proportional to the unit matrix $I_2 \otimes I_4$.  The Zeeman effect of the magnetic field is thus to shift both zero-energy states equally, so that both are above, or below, the Fermi level \cite{remark1}. From Eq. (4) we see that it is the OP belonging to the $U(1)$  factor that develops an expectation value in this case, which is precisely the HKM state  again.

 Within the $\Phi$-representation it is evident that the above result holds irrespectively of the symmetry of the superconducting OP. For each one of the four possible choices the $U(1)$ part of the $U(1)\times Cl(3)$ algebra of the core states will always be the HKM, since the matrix $i \gamma_1\gamma_2$ anticommutes with all the superconducting states, while at the same time it commutes with all the insulating states, which are the exclusive members of the $Cl(3)$. Since the Zeeman term is proportional to the unity operator only the {\it universal} $U(1)$ part of the algebra develops the expectation value for any symmetry of the OP.

  There is {\it always} the $U(1)\times Cl(3)$ algebra of possible OPs in the core, irrespectively of the type of order supporting the vortex. For example, in the $\Psi$-representation a vortex in the N\' eel OP in the x-y plane, $\sim ( \sigma_1 \cos\theta  + \sigma_2 \sin\theta ) \otimes i \gamma_1 \gamma_2 $, in the core may exhibit the p-wave superconductor, the CDW, and the z-component of the same N\' eel OP ($\sigma_3 \otimes i\gamma_1 \gamma_2$) \cite{herbutAF}. The most general Dirac Hamiltonian with a vortex can be written as
   \begin{equation}
   H_{gen} = M_1 \hat{p}_1 + M_2 \hat{p}_2 + o_1 M_3 +  o_2 M_4,
   \end{equation}
   where $(o_1, o_2) = |o(r)| (\cos\theta, \sin\theta) $ is the OP, and $\{ M_i \} $ are four arbitrary eight-dimensional Hermitian matrices satisfying $[M_i, M_j]_+ =2 \delta_{ij}$. Since $\{ M_i \}$ form a representation of the Clifford algebra $Cl(4)$, by a unitary  transformation it can be reduced to a block-diagonal form $M_i = \alpha_i \oplus \beta_i$, where $\alpha_i$ and $\beta_i$ are four-dimensional \cite{schweber}. Since $\{ \alpha_i \}$ and $\{ \beta_i \}$ then provide two (in general, different) Hermitian four-dimensional representations  of $Cl(4)$, they are equivalent \cite{schweber}. Any $H_{gen}$ may therefore be transformed into the block-diagonal form of $H_{BdG}$ as given in Eq. (3), with  {\it the same} four-dimensional matrix operator in both blocks. The only four operators that anticommute with such a Hamiltonian are obviously $I_2 \otimes \gamma_0$ and $\vec{\sigma} \otimes \gamma_0$, which form the algebra $U(1)\times Cl(3)$.

If the superconducting OP in graphene is induced by a layer of a type-II superconductor laid on top, the Zeeman energy  would roughly be $\epsilon_z \approx H\times$kelvin, if the magnetic field $H$ is expressed in tesla. The leading effect of the graphene lattice, for example, comes from the $\sim 2\vec{K}$ Fourier component of the vortex configuration, which in $\Phi$-representations is proportional to the matrix $\sigma_1 \otimes I_4$. It is straightforward to check using our formalism that this perturbation favors the scalar BDW in the core, in agreement with the result of direct lattice calculation \cite{ghaemi}. The magnitude of this splitting should crudely be $\epsilon_{dw} \sim |\Delta| (a/\xi) \approx (10^2 a/\xi)^2$kelvin, where $a\approx 1 \AA$ is the lattice spacing. In the $Nb_3 Ge$ with $T_c\approx 23 K$ for example, $\epsilon_{dw} \approx 1 K$. The upper critical field, however, is $\sim 40$ Tesla, so at the magnetic fields of few tesla the order in the core should predominantly be the HKM. It is also possible to increase the HKM component of the OP by adding a component to the magnetic field parallel to the superconducting layer so not to suppress the condensate. A distinct sign of the HKM OP in the core would then be the {\it increase} of the gap in the local density of states with the total magnetic field at a finite temperature. This follows from the generalization of Eq. (4) to finite temperatures:
\begin{equation}
\langle OP \rangle = \frac{\psi_0 ^\dagger \psi_0  }{2} (\tanh( \frac{\epsilon_z+\epsilon_{dw}}{2 k_B T} ) + s \tanh( \frac{\epsilon_z-\epsilon_{dw}}{2 k_B T} )),
\end{equation}
with $s=1$ for the HKM, and $s=-1$ for a linear combination of the CDW and BDW. Assuming $\epsilon_z$ and $\epsilon_{dw}$ to differ by a factor of 5 or more one OP is by an order of magnitude larger than the other over a wide range of temperatures. If the dominant order is any density wave, for $\epsilon_{dw} \gg \epsilon_z$,  the OP would in contrast {\it decrease} with the magnetic field.

It may also be useful to note that the competing orders behave differently under the changes of the signs of vorticity and/or magnetic field. For example, if the sign of both the  vorticity and the magnetic field is reversed the BDW changes into its complementary pattern \cite{ghaemi} whereas the HKM state remains invariant. A probe sensitive to the sign of the gap would thus help distinguish between the different core states.

This work was supported by NSERC of Canada. The author wishes to thank V. Juri\v ci\' c, B. Roy, B. Seradjeh, and P. Nikoli\' c for discussions, and to the Aspen Center for Physics where a part of this work was carried out.


\begin{thebibliography}{99}
\bibitem{remark1} Since the zero-energy  states are the Majorana fermions \cite{wilczek} such a shift does not change the particle number.
\end{thebibliography}
\end{document}